\documentclass[10pt,conference]{IEEEtran}
\usepackage{cite}
\usepackage{amsmath,amssymb,amsfonts}
\usepackage{algorithmic}
\usepackage{tabularx}
\usepackage{graphicx}
\usepackage{textcomp}
\usepackage{xcolor}
\def\BibTeX{{\rm B\kern-.05em{\sc i\kern-.025em b}\kern-.08em
    T\kern-.1667em\lower.7ex\hbox{E}\kern-.125emX}}
\begin{document}

\title{Applications of Causality and Causal Inference in Software Engineering\\
}

\author{\IEEEauthorblockN{Patrick Chadbourne}
\IEEEauthorblockA{\textit{Department of Computer Science} \\
\textit{Boise State University}\\
Boise, ID, USA \\
patrickchadbourn@u.boisestate.edu}
\and
\IEEEauthorblockN{Nasir U. Eisty}
\IEEEauthorblockA{\textit{Department of Computer Science} \\
\textit{Boise State University}\\
Boise, ID, USA \\
nasireisty@boisestate.edu}
}

\maketitle

\begin{abstract}
Causal inference is a study of causal relationships between events and the statistical study of inferring these relationships through interventions and other statistical techniques. Causal reasoning is any line of work toward determining causal relationships, including causal inference. This paper explores the relationship between causal reasoning and various fields of software engineering.
This paper aims to uncover which software engineering fields are currently benefiting from the study of causal inference and causal reasoning, as well as which aspects of various problems are best addressed using this methodology. With this information, this paper also aims to find future subjects and fields that would benefit from this form of reasoning and to provide that information to future researchers.
This paper follows a systematic literature review, including; the formulation of a search query, inclusion and exclusion criteria of the search results, clarifying questions answered by the found literature, and synthesizing the results from the literature review.
Through close examination of the 45 found papers relevant to the research questions, it was revealed that the majority of causal reasoning as related to software engineering is related to testing through root cause localization. Furthermore, most causal reasoning is done informally through an exploratory process of forming a Causality Graph as opposed to strict statistical analysis or introduction of interventions. Finally, causal reasoning is also used as a justification for many tools intended to make the software more human-readable by providing additional causal information to logging processes or modeling languages.

\end{abstract}


\section{Introduction}
Causal inference is an up-and-coming field of study that is applied and advanced in several scientific research areas for various disciplines (e.g., Medicine, Statistics). Despite the growth of popularity in this field of research, it is relatively underutilized in software engineering despite its significant potential for applicability to numerous common problems in the field. This paper focuses on providing an empirical study of causal inference and its usage in software engineering, as well as identifying areas of opportunities for its application through a systematic literature review. The literature review will consist of discovering and collecting research papers relating both generally to causal reasoning and causal inference techniques and specifically to various problems and fields of study in software engineering. The collection of these papers will be used to answer several questions relating to the specific applications of causal reasoning and software engineering techniques. The answers will then be used to fuel the synthesis of results.

This study seeks the answers to two primary questions and their numerous related queries:

\emph{RQ1: Which fields of software engineering currently benefit the most from causal inference?}
The primary objective of this project is to discover which fields and areas under the broad Software Engineering umbrella are currently gaining the most benefit from causal inference and causal reasoning. The authors hypothesize that through an improved understanding of which fields are currently finding success using these techniques, it will be possible to find under-served areas of Software Engineering which may benefit from the application of new techniques, as well as areas that are already well-suited to using these techniques and may further benefit from additional research done regarding similar methodologies. 

%By gathering and examining research papers relating to software engineering that include mention of causal inference or causal reasoning, this study will examine not only the field within software engineering that is specifically leveraging these causal inference techniques but will also be able to investigate which problems within that field are finding benefits from the statistical analysis provided by various forms of causal reasoning. 

By answering this question, we will also be able to uncover the inverse, which relates to which fields of software engineering are currently under-served by applying causal inference techniques. Finding this information will be of undoubted value to future researchers looking for relevant fields to apply their knowledge of causal reasoning by providing not only these under-served fields but also the relevant types of problems that have been solved in other fields using similar techniques.

%Furthermore, by discovering which fields of software engineering utilize these forms of reasoning, it will be possible to draw comparisons between them to find out how causal inference utilized in a certain field differs significantly from applying those same techniques in a separate field. Finally, beyond simple differences in the application of techniques, it is also of interest whether similar techniques are met with differing quality of results depending upon which field of software engineering they are applied to.

\emph{RQ2: Which causal inference techniques are being applied to software engineering?}
The secondary objective of this undertaking is to outline the currently used causal inference techniques and their effectiveness to uncover which techniques are applied more successfully or to determine if any underlying trends provide evidence towards certain techniques being more applicable or beneficial for certain classifications of problems.

The authors hope to find patterns that can be used to form recommendations for future researchers and practitioners to better plan and implement their usage of causal inference techniques, with additional knowledge regarding what software engineering research has previously benefitted from such techniques, as well as which areas of software engineering are likely to help despite the relative lack of previous work combining the field and methodology.

%While gathering data through the examined research papers, this study will also categorize the applied techniques for causal inference. Through this categorization, the study will provide a deeper understanding of the potential limits of these techniques and relevant benefits or drawbacks to differing methodologies.

%Through comparisons of the applications of the various techniques found, this study will seek to conclude when multiple techniques should be used over other competing possibilities and to highlight the relative success rate of any individual methodologies over the others.

%It is also potentially possible that through performing this systematic literature review, there will be specific strategies related to causal reasoning that are used as part of successful software engineering research that is unused or under-utilized by any of the other available software engineering research and through uncovering these techniques in the light of comparison it may better demonstrate the value in applying these specific strategies to other problems within different fields of software engineering.

\section{Related Works}

The information in this section relates to the foundation of knowledge upon which this paper has been written, both relating to the topics discussed within and the methodology for writing the paper itself.  
%Causal inference is its own niche relating to statistical analysis and there is a significant number of related papers to the various techniques that will hopefully be uncovered by this systematic literature review. Software engineering is an extremely broad field and there are no shortage of relevant research papers that can be looked at, especially relating to various methodologies and strategies for performing software engineering using specific forms of analysis such as are done in causal inference. Additionally papers have been looked at relating to the steps taken to perform a systematic literature review.

It is important to our understanding of how causal inference techniques are being applied to software engineering that we have a robust knowledge regarding how those techniques are intended to be used and other related techniques that could be used in their place. Pearl~\cite{b1} has written the primary resource cited by papers in this field: a book underlying the majority of thinking related to causal inference as a field of study. Wong~\cite{b2} extends this thinking into the area of computation and outlines a methodology to apply causal inference techniques to large sets of software-driven data. Although differing in the approach, a pre-existing study by Kitchenham et al.~\cite{b3} looked into the possibility of creating an evidence-based software engineering model, which also made use of causal reasoning to provide causal proof of the efficacy of their suggested changes.

Baah et al.~\cite{b4} applied causal reasoning methodologies to perform observational studies on software fault localization on test outcomes and profiles. On the other hand, Küçük et al.~\cite{b5} examined this fault localization process and improved it with different causal inference techniques. In addition, Arya et al.~\cite{b6} discussed applying causation-based strategies to the field of artificial intelligence for information technologies. By closely examining these studies, we can see the strategy these researchers employed to find their results and what fields of software engineering they applied these causal inference techniques.

\begin{figure*}[h!t]
\centerline{\includegraphics[width=\textwidth]{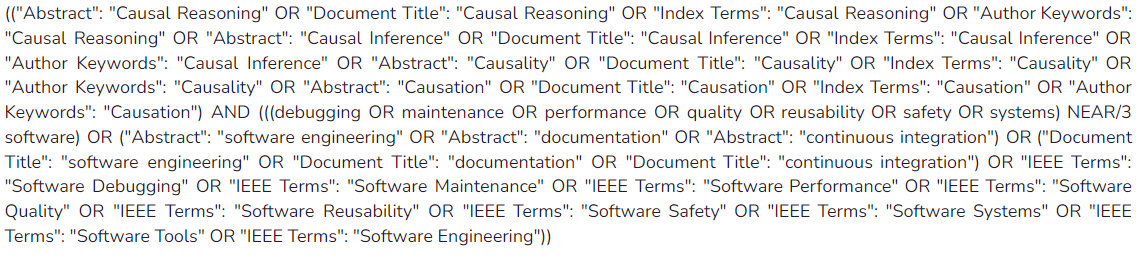}}
\caption{The full search query used for the source paper collection.}
\label{fig}
\end{figure*}

%\subsection{Second Area of Research}
%\emph{RQ2: Which causal inference techniques are being applied to software engineering?}

%During the gathering of data through the examined research papers this study will also categorize the techniques relating to causal inference that are applied. Through this categorization the study will provide deeper understanding of the potential limits of these techniques as well as relevant benefits or drawbacks to differing methodologies.

%Through comparisons of the applications of the various techniques that are found this study will seek to draw conclusions about when various techniques should be used over other competing possibilities as well as to highlight the relative rate of success of any individual methodologies over the others.

%It is also potentially possible that through performing this systematic literature review there will be specific strategies related to causal reasoning that are used as part of successful software engineering research that are unused or under-utilized by any of the other available software engineering research and through uncovering these techniques in the light of comparison it may better demonstrate the value in applying these specific strategies to other problems within different fields of software engineering.

\section{Methodology}

This section details the specific tools and strategies that were used to gather the primary source research papers, the criteria for whether a paper was included or excluded from the study, as well as several intermediate questions that were answered on a per-paper basis to gather the necessary data from which our results would be synthesized. Including this information is provided both to give insight into the methodology used and to provide detailed steps to facilitate the reproduction of this study for future researchers or peer reviewers. The details included should also provide insight into the reasoning behind various decisions made as part of the research process and offer justification for those decisions as well.

\subsection{Search Query}

This study was performed using results collected from the IEEE Xplore research database using the search query that can be seen in Fig~\ref{fig}. Although the query is complex to include as many specifically relevant papers as possible while excluding those similarly unrelated, the intention of the query is quite simple. The query combines two basic concepts and requires that both of those concepts be present in a paper for it to be returned in the search while using several `OR' operators to adequately cover the full extent of the concept while not using an overly broad term that might include undesired results.

The first of these core concepts within the search query is a causal inference or causal reasoning. As the study of causal inference can be done in such a way that is purely about statistical analysis with no relation to software engineering, we searched for various terms such as ``Causality" or ``Causation" in a paper's abstract, title, and index terms as these were the places where they would be most likely to show up while relating to being used as a methodology for researching software engineering.

The second core concept was that of Software Engineering, as the goal was not to find specific subjects within software engineering. This portion of the query comprised either finding specific terms in conjunction with ``software" in a document's title or using the `IEEE Terms' category search.

In making use of the `IEEE Terms' categories, the query specifically utilized the categories listed as sub-categories of software engineering in the IEEE Taxonomy. Once again, the broader IEEE category of ``Software Engineering" was not used as a catch-all term to exclude results that would be broader than desired. Despite seeming like such a category would be highly relevant to the results desired by this study, the inclusion of the entire Software Engineering category as described by IEEE would extend beyond the scope of this study's research.

Certain terms needed to be avoided while constructing this query despite their relevance to the desired search results due to overlapping definitions within other fields, and although it may have been possible to construct a more complicated query that would then exclude the undesired results it was decidedly easier to simply not include the problematic search terms in the first place. These terms included words such as ``integration" which may be related to either the software engineering concept of continuous integration, or various statistical analysis techniques, however the term is used far too broadly in too many fields to consistently return results relevant to this source material. 

This search query returned 191 results after performing several iterations to refine and tune the results, minimizing irrelevant results while not losing relevant papers and ensuring that the papers adhered to a strict set of criteria.

\subsection{Inclusion and Exclusion}

Due to the restraints of the resources available to this study, it was important to ensure that the results returned were as relevant as possible to reduce the manual exclusion required to receive the final selection of pertinent source papers. This included automatically excluding any search results that did not match the general criteria of the study. For example, papers were required to be written in English, that the paper is a complete study and not an artifact, and that the paper is published in a conference or journal. With these criteria in place, our search yielded 191 results. Of the original 191 results returned from the query, 45 papers were selected after a two-part manual review.

The primary step of the review process was to briefly read each abstract of the initial 191 papers and mark it as either `Definitely Relevant', `Potentially Relevant', or `Not Relevant' to the study. This was done by ensuring that the paper was strictly related to a software engineering concept and that the paper's authors, in some way, used causal inference or causal reasoning to address the problem. Although it was trivially easy to ensure that a paper addressed a software engineering concept, it wasn't easy to ensure that their usage of causal reasoning was relevant to the study of causal inference that is of interest to this study. To ensure that papers were not erroneously excluded and to capture the full spectrum of research utilizing various types of causal reasoning, almost any usage of causal reasoning was included, no matter how minor of a role it played. Later categorizing the papers by the extent of their usage of causal reasoning would therefore be essential to justify the inclusion of this full extent of papers.

After determining a paper's relevance, the results marked as `Definitely Relevant' and `Potentially Relevant' were closely examined to answer four intermediate questions relating to the study. If any paper could not answer the first three questions, it was excluded from consideration for not having information related to the study. The fourth question was optional as it only related to the quality of their results, where some papers performed valuable research that was relevant to both software engineering and causal reasoning but did not result in data that could be validated. Of the initial 191, the papers that could answer the intermediate questions were kept and formed the 45 primary sources upon which the results are based.

\subsection{Intermediate Questions}

In order to answer the original two research questions, it is necessary to ask some intermediate questions of each paper in the primary sources. These intermediate questions are directly related to the specifics of that individual paper. They are intended so that the answers to the questions can be aggregated to shed light on the answers to the original research questions. These intermediate questions are as follows:

\begin{itemize}
    \item \emph{What field of software engineering does the paper relate to?}
    \item \emph{What is the problem within this field that it addresses?}
    \item \emph{How is causal reasoning being applied to the problem?}
    \item \emph{Was the application of causal reasoning valuable to their solution?}
\end{itemize}

For each question a number of categories was created to classify the answers for each paper. For the fields of Software Engineering this is limited to the phases of the software development life-cycle, with categories for design, development, testing, deployment, maintenance, and collaboration.

Categories created for the second question are unique to each stage in the software development life-cycle that is selected as an answer to the first question. They are far more specific to the individual problems addressed in each paper. The categories are somewhat generalized where possible so that similar or different problems in the same area are grouped. This prevents each paper from having its unique category for the unique variation of a problem it addresses.

The answers to the third question are again generalized, grouping the majority of results into the closest fitting categorization for the type of causal reasoning the authors applied. For a significant number of papers, there was no clear ``technique." Still, a general causality-focused approach to their study methodology. For all primary sources where this was the case, they were placed in the ``Groundwork to Facilitate Causality" group about the benefits the authors were trying to gain via their inclusion of causal reasoning.

Finally, the answers to the fourth question reflect the quality of the paper's results and are grouped into four categories: ``Outperformed Expectations", ``Performed Adequately", ``Provided Mixed Results", and ``Underperformed". These answers were based on the authors' measurement of success, as there was no feasible way to manually grade the outcomes of each project without an unrealistic amount of intensive labor.  

\subsection{Data Aggregation and Synthesis}

Once the results are gathered and recorded, the final stage of the study is to take those results and examine them to locate trends and draw conclusions. With results from each segment of the software development life-cycle grouped, it was possible to compare the relative performance and popularity of causal inference techniques.

Additionally, conclusions were drawn from examining the frequency of various problems within the groupings and the relative frequency of different techniques to papers addressing similar problems. The goal was to determine if researchers approaching similar tasks with different methodologies would find notably different results and, if so, if those differences can be generalized to broader categorizations of software engineering fields.

\begin{table}[htbp]
\caption{Papers by Software Engineering Field}
\begin{center}
\begin{tabularx}{\linewidth}{X|c}
\hline
\multicolumn{2}{|l|}{\textbf{Software Engineering Fields}} \\
\hline
Testing & 17 \\
\hline
Deployment & 9 \\
\hline
Design & 7 \\
\hline
Development & 5 \\
\hline
Maintenance & 5 \\
\hline
Collaboration & 1 \\
\hline
\end{tabularx}
\label{tab1}
\end{center}
\end{table}

\section{Results}

This section discusses the results of answering the intermediate questions for each primary source paper. It examines trends in the resulting data and where trends were expected but not found.

\subsection{Software Engineering Fields}
The broadest results can be seen in Table I, which shows that most of our papers were related to testing. For the purposes of this paper, anything about fixing code defects, generating tests, improving tests, or otherwise interacting with the process of discovering and repairing broken code was considered part of testing. Maintenance and deployment as categories had some overlap with testing due to their relevance to software defects. However, papers in these categories did not handle software correctness or fixing. Instead, they dealt with logging errors when they occurred, either as part of the software development life-cycle or during the deployment process. The single paper in the collaboration category was specifically about discovering the cause of changes in user metrics while collaborating on open-source repositories and was somewhat of an outlier among the other papers.

\subsubsection{Testing}

With 11 results, most of the papers related to testing dealt with \emph{fault localization}~\cite{b43,b44,b45,b46,b47,b52}; the process of discovering the root cause of a known bug. This was a clear fit for causal reasoning as at any point when an engineer is faced with a known error in the code, one of the primary steps in resolving it is to discover which piece of code is resulting in the error. This can sometimes be trivially simple, as a compiler or automated analysis will show precisely where the error is. Still, it can be quite complex in some situations as the root cause will be far removed from where the error appears to occur. Since this cause-effect nature is integral to software debugging, it is unremarkable that it would have so many applications for causal reasoning.

An additional three papers \cite{b25,b30,b39} dealt with predicting code defects by examining software, determining where interventions would likely cause errors, and using the inference to examine where errors were already likely to exist.

\subsubsection{Deployment}

Similarly to the results for the testing papers, the deployment papers have a clear majority at 6 papers relating to \emph{software monitoring}~\cite{b15,b20,b22,b29,b36,b40}. These papers often presented novel techniques for using a causal reasoning framework to augment existing software monitoring processes to provide additional causal information to the outputted data. The reasoning for this is likely very similar to that for fault localization. When monitoring software, anything that would warrant notice would only benefit from providing the end user with additional information about what had caused the output.

\subsubsection{Design}

The papers in the design category were probably the most unique of those collected in the primary sources. However, the drawback of having the most tenuous connections to causal reasoning is. A significant number of these papers related to the creation of new, or modification of existing, modeling languages that placed a more significant emphasis on cause-effect relationships in the software they modelled~\cite{b9,b10,b16}, following more of a literal meaning of causation as opposed to the more statistical analysis methodology. However, some papers did stand out from this trend with attempts at automated requirement analysis~\cite{b13,b48} that made use of causal interventions to determine which requirements would necessitate significant changes to the design of the software.

\begin{table}[htbp]
\caption{Papers by Causal Reasoning Application}
\begin{center}
\begin{tabularx}{\linewidth}{X|c}
\hline
\multicolumn{2}{|l|}{\textbf{Causal Reasoning Applications}} \\
\hline
Causality Graph & 17 \\
\hline
Groundwork to Facilitate Causality & 12 \\
\hline
Bayesian Model & 5 \\
\hline
Granger Causality Test & 3 \\
\hline
Difference-in-Differences Model & 2 \\
\hline
Counterfactual Prediction & 2 \\
\hline
Statistical Analysis & 1 \\
\hline
Evidential Network & 1 \\
\hline
Propensity Model & 1 \\
\hline
\end{tabularx}
\label{tab1}
\end{center}
\end{table}

\subsection{Causal Reasoning Applications}
Table II summarizes all gathered papers by the methodology in which causal reasoning was applied. Of these results, the two most common categories of the \emph{Causality Graph} and the \emph{Groundwork to Facilitate Causality} were undoubtedly the most generals. However, separating these categories into more specific groupings would not have been feasible as they both included the most general methodologies in qualifying as causal reasoning. Unlike something more specific such as intervention through a \emph{Bayesian Model} or demonstration of a cause-effect relationship with a \emph{Granger Causality Test} where the applied technique is straightforward in its motivation and implementation, these more general categories were always specifically implemented to exactly meet the requirements of the problem being addressed by the paper.

\subsubsection{Causality Graphs}

The concept of a causality graph itself is not a strict one but instead refers to the general workflow process that many of these papers~\cite{b43,b44,b45,b31,b18,b19,b20,b21,b22} followed where the problem was broken down into a graphical form. The connections between various events or parameters were illustrated and explored to determine which areas were more likely to have causal relationships. Although no statistical analysis was done relating to this causality in these papers, it emphasized the importance of causation in their work. It suggested that their problems might be well-suited to a solution that would use a more strict application of causal inference in the future.

\subsubsection{Groundwork to Facilitate Causality}

Papers following a groundwork to facilitate causality~\cite{b10,b12,b16,b29,b36,b37,b38} were motivated by a desire to find causal relationships in the problem space they were exploring but did not use any meaningful or distinct techniques that could be recorded. Instead, these papers focused on understanding what these causal relationships could look like relating to their problem and began to do work that would better uncover these relationships so that future research could have a better starting point to do statistical analysis or intervention techniques.

\subsubsection{Bayesian Model}

Of the distinct and notable techniques leveraged to perform causal inference on the various research problems in software engineering, the most notable were the papers that leveraged Bayesian Models~\cite{b26,b27,b28,b39,b51}. These papers applied a methodology where training data is passed into a Bayesian Model, and a predictor is leveraged to analyze the effects of changes to this data. Through creating a controlled environment where the experimenter can ensure that nothing other than the intended change is made, it is possible to make a strong claim of causality between that change and the ultimate effects on the outcome data. 

\section{Discussion and Conclusion}

Although an enlightening exploration of a significant number of papers relating to software engineering and causal reasoning, it was unfortunately difficult to draw any strongly supported conclusions about applications of individual causal reasoning techniques due to the minimal number of papers that made use of them as such. Looking at more general trends showed that this is undoubtedly a field with great potential, as many papers professed significant results while using these techniques. A number of the proposed automated tools were undoubtedly for problems that will continue to persist for some time, such as software debugging, monitoring, and logging. Additionally, some of the more specific research papers handling things such as automated requirement synthesis made solid arguments for why applications of causal reasoning are a strong candidate for additional research and attention.

This systematic literature review should provide a solid foundation for future researchers interested in future software engineering trends or those interested in causal inference as a field. Furthermore, it should continue to be a strong fit as a number of the causal inference techniques that are growing in popularity and becoming more beneficial to several fields are performed using machine learning and data science techniques, both fields which have a great deal of overlap with software engineering already.

Unfortunately, some aspects of the review were less valuable than others, such as the results relating to the fourth intermediate question about whether the application of causal reasoning was beneficial to their results. As this question could only feasibly be answered by examining what the authors claimed about their work, it was subject to a great deal of bias. Then our interpretation of those claims undoubtedly introduced an additional layer of prejudice to the equation. This was further compounded by the fact that not all papers reported the quality of their results, some merely leaving their research with the expectation that it was simply a beginning and that actual results would not exist until it was taken further, either in a future study by the authors or in a continuation by other researchers.

Regarding this systematic literature review, there is a great deal of room for future work, either by broadening the scope of the initial search and discovering more papers with concrete applications of causal reasoning or by doing a more in-depth examination of the individual causal reasoning techniques to form a sort of primer for applying these techniques to software engineering disciplines.

\end{document}